\documentstyle[aps,multicol,prl,epsfig]{revtex}
\begin{document}
\title{\bf\boldmath Charge and orbital order in RNiO$_3$ (R=Nd,Y) by LSDA+U method}
\author{Susumu Yamamoto and Takeo Fujiwara}
\address{Department of Applied Physics,  University of Tokyo,
Tokyo 113-8656, Japan}
\date{\today}

\draft
\maketitle
%-----------------------------------------------------------------%
\begin{abstract}
Electronic structures of NdNiO$_3$ and YNiO$_3$ are calculated by
using LSDA+U method  with rotational invariance. 
The Jahn-Teller distortion is not allowed under the observed magnetic ordering. 
No orbital order on Ni sites cannot be observed in the 
calculation of  both systems but different types of charge ordering. 
In a small distorting system NdNiO$_3$,
all Ni ions are trivalent  and  oxygen sites 	have a particular
ordering, both in charge and orbital.  
In a large distorting system  YNiO$_3$, the charge disproportionation 
occurs,  2Ni$^{3+} \rightarrow$ Ni$^{2+}+$Ni$^{4+}$.  
Therefore, the charge ordering stabilizes 
the asymmetry of the arrangement of Ni magnetic moments in both systems. 
\end{abstract}
%-----------------------------------------------------------------%
\begin{multicols}{2}
\narrowtext
%\section{Introduction}%
Perovskite transition metal oxides are now of high interest because of
the large variety and the possible controllability of their physical
properties. 
Perovskite nickelates RNiO$_3$ (R = an
trivalent rare earth or Y ions) can be classified into three different 
categories, 
according to their tolerance factor ($t=d_{\rm R-O}/\sqrt{2}d_{\rm
Ni-O}$).~\cite{RNiO3-N-DIFF,RNiO3-Mag-DIFF,RNiO3-Mag-DIFF2,YNiO3,RT-CO}
One is those whose tolerance factor is much smaller ($t<0.91$) 
and has a larger distortion, such as LuNiO$_3$ or YNiO$_3$. 
They are antiferromagnetic insulators at low temperatures.
This class of RNiO$_3$ undergoes transition to
paramagnetic insulator at the Neel temperature $T_{\rm N}$ and 
also has another phase transition from paramagnetic insulator to 
paramagnetic metal (M-I transition) at very high  transition temperature.  
Second is those whose tolerance factor is intermediate 
($0.91 < t < 0.93$), such as NdNiO$_3$ or PrNiO$_3$. 
These materials are also antiferromagnetic insulators at
low temperatures and above $T_{\rm N}$ they are
paramagnetic metals.~\cite{NdNiO3-el}
Third is LaNiO$_3$ where $t=0.94$ and is paramagnetic metal.

Low temperature phase of above mentioned first and second classes of
RNiO$_3$ have the unique magnetic 
structures.~\cite{RNiO3-N-DIFF,RNiO3-Mag-DIFF,RNiO3-Mag-DIFF2,YNiO3} 
Its magnetic diffraction peak is characterized by the propagation vector
$\vec{k}=(1/2,0,1/2)$ and the magnetic unit cell is
identified as the $2\times 1 \times 2$ supercell of the
crystallographic  lattice.~\cite{RNiO3-N-DIFF,YNiO3}
The magnetic order is specified to be alternating
ferromagnetic (FM) and antiferromagnetic (AFM) couplings  $\rm
\cdots Ni^\uparrow -\stackrel{\mbox{\tiny FM}}{O}-Ni^\uparrow -
\stackrel{\mbox{\tiny AFM}}{O}-Ni^\downarrow -\stackrel{\mbox{\tiny
FM}}{O}-Ni^\downarrow -\stackrel{\mbox{\tiny AFM}}{O}-Ni^\uparrow -
\stackrel{\mbox{\tiny FM}}{O}-Ni^\uparrow \cdots$ along three
crystallographic directions.

On the contrary to the asymmetric
arrangement of magnetic bonds (FM/AFM) around each nickel site, Ni-O
distances around each Ni ion are almost equal to each other. 
For instance, the difference between the longest and the shortest Ni-O
bonds in one NiO$_6$ octahedron is at most 2\%
in RNiO$_3$.~\cite{RNiO3-N-DIFF,RNiO3-Mag-DIFF2,RT-CO} 
%as listed in Table~\ref{tble-bondL}
The Jahn-Teller distortion is absent in RNiO$_3$, 
even in the largely distorted system YNiO$_3$, 
whose tolerance factor is 0.88.
Experimentally there is only one crystallographic site
for Ni ions in NdNiO$_3$ and  two different sites for Ni ions in
YNiO$_3$.
Therefore, one can expect trivalent ions Ni$^{3+}$ in NdNiO$_3$. 
On the contrary, Ni ions in YNiO$_3$ are divalent and quadrivalent, {\it i.e.}
the charge disproportionation 2Ni$^{3+} \rightarrow$
Ni$^{2+}+$Ni$^{4+}$ occurs.

%-----------------------------------------------------------------%
%\section{Methods used in the calculations}%
In this letter, we study the electronic properties of the
AFM insulating phase of two perovskite nickelates, NdNiO$_3$ and
YNiO$_3$, which are typical ones of above mentioned two respective
classes. 
We use the LSDA+U method with the rotational invariance
in conjunction with the LMTO-ASA method.~\cite{LSDA+U-3} 
The LSDA+U method counts
the electron-electron interaction between localized orbitals by the
Hartree-Fock type interaction term.  
The ionic
positions and lattice parameters used in the present calculations are
imported from the diffraction experiments.~\cite{RNiO3-N-DIFF,YNiO3}
The unit cell here is the $2 \times 1 \times 2$ crystallographic
supercell containing 16 Nd or Y ions, 16 Ni ions, and 48 O ions 
and, in addition,  32 empty spheres, totally 112 atomic spheres. 
Sixty-four k-points in the Brillouin zone are sampled in the 
calculation of the density of states. 
NdNiO$_3$ is  antiferromagnetic up to $T_{\rm N}=200{\rm K}$, 
but the Nd spin moment vanishes at about
30K.~\cite{RNiO3-Mag-DIFF} 
Therefore, three 4f electrons of Nd ion are
counted in the frozen core.  
We also calculated the electronic structure of NdNiO$_3$ with 4f electrons 
in  valence states, and the results show no significant difference. 
The Coulomb and exchange parameters $U$ and $J$ of Ni ions are 
fixed to be 7.0~eV and 0.88~eV, respectively, 
through all the LSDA+U calculations. 
These values are consistent with photoemission experiments and the 
results of the LSDA calculations.~\cite{valuesU&J,Anisimov}
Calculated results are not changed much over a large range of
values of $U$ and $J$.

The crystallographic space groups 
of NdNiO$_3$ and  YNiO$_3$  are orthorhombic  $Pbnm$ 
and monoclinic $P2_1/n$, respectively.
The group theoretical analysis shows that, using the projection
operator method, the Jahn-Teller distortion of one NiO$_6$ octahedron 
can not be transfered  over the whole enlarged $ 2 \times 1 \times 2$ 
supercell with ${\vec k}=(1/2,0,1/2)$.~\cite{RNiO3-GroupT}
Therefore, the Jahn-Teller distortion actually 
cannot appear in the observed spin ordered state.

%-----------------------------------------------------------------%
%\section{Results and discussion}%
NdNiO$_3$: 	
There are two possible spin configurations in ${\rm NdNiO_3}$ satisfying the observed 
transfer vectors ${\vec k}=(1/2,0,1/2)$. One is that  the spin 
magnetic moments align ferromagnetically on the (101) plane  and 
the planes are stacked with a doubled period as 
$\uparrow\uparrow\downarrow\downarrow\uparrow\uparrow\downarrow\downarrow\cdots\cdots$, 
whose magnetic space group is monoclinic $P_ba$.~\cite{ML}
This spin configuration is assumed in the present letter. 
Another possible spin configuration with
$\uparrow\uparrow\downarrow\downarrow$ spin order could be the 
doubled period checkerboard stacking along crystallographic $y$-axis, 
and the magnetic space group is orthorhombic  $C_amc2_1$.~\cite{ML}
The state of $C_amc2_1$ shows  orbital ordering 
of e$_g$ states on Ni sites, 
but the resultant calculated total energy 
is higher by 0.18~eV per $2\times 1\times 2$ cell than that of 
$P_ba$.  Because  there is no possibility of 
 subsidiary  Jahn-Teller distortion to lower the total energy, 
$C_amc2_1$ could not be the symmetry to be considered.
The total energy is also calculated in a fictitious antiferromagnetic
state of $P{b^\prime}nm^\prime$ whose  magnetic unit cell is identical to
the crystallographic one  and is higher by 2.0~eV per $1 \times 1\times
1$ cell.  
%---------------------
\begin{figure}[htbp]
\centering
\vspace{0.5cm}
	\caption{Projected
	density of states of trivalent nickel Ni$^{3+}$ in NdNiO$_3$.
The energy zeroth is set at the top of the occupied states $E_{\rm
F}$. 
The local $z$ axis directs from Ni to one of oxygen atoms with
the shortest Ni-O distance, and the local $x$ and $y$ axes to the
appropriate O atoms.}
\label{Fig-NdNiO3-PDOS}
\end{figure}
%---------------------

Figure \ref{Fig-NdNiO3-PDOS} shows the projected density of states on
the Ni 3d orbitals in NdNiO$_3$. 
The system is insulator with a gap $E_g=0.11~$eV. 
Each nickel ion has the local magnetic moment $\pm 1.1\mu_{B}$ 
within the atomic sphere of a radius $2.51~{\rm au}$,
contrary to  the observed value of $\pm 0.9\mu_B$.~\cite{RNiO3-Mag-DIFF}
It should be noticed that the local magnetic moment cannot be uniquely
defined. 
Furthermore, NdNiO$_3$ is not a simple antiferromagnetic insulator but 
a dynamical effect is essential, which may be an origin of this 
discrepancy.~\cite{NdNiO3-el}
There is no distinctive variation in partial spin density of
states of Ni site in each magnetic sublattice. 
An e$_g$ band lies on the energy range $E_{\rm F}-0.6~{\rm eV}
\sim E_{\rm F}+1.2~{\rm eV}$. 
The numbers of states in the energy ranges
 $E_{\rm F}-0.6~{\rm eV} \sim  E_{\rm F}$ and $E_{\rm F}+0.1~{\rm eV} \sim 
 E_{\rm F}+1.2~{\rm eV}$ are 16, respectively, which are the 
one occupied and one vacant e$_g$ states per  Ni ion.
Actually, these e$_g$ orbitals extend
over surrounding oxygen sites from Ni ions due to strong hybridization
between Ni e$_g$ and O p orbitals. 
The extended occupied e$_g$ orbital has
a 60\% weight on p orbitals on surrounding six oxygens, 
a 10\% weight on an individual oxygen.  
Therefore, one would establish a model where 
one occupied  e$_g$ state with majority spin locates 
at the top of the valence bands, 
and it  hybridizes strongly with the p states on nearby O ions. 
This is the molecular orbital  $\sigma^*$ state.~\cite{Sugano-Shulman}
Then  one can assign  all nickel ions in NdNiO$_3$ 
to be  trivalent Ni$^{3+}$ (t$_{2g}^6$e$_g^1$), even though 
the Ni ion is not truly ionized by +3 charge.  
In the projected density of states of Ni ion site, one observes 
 a large amount of e$_g$ states  at the bottom of the d bands 
, which  are the  bonding states  between Ni d and O p,
corresponding  to  the  $\sigma$ states 
in the molecular orbital picture.~\cite{Sugano-Shulman}

%----------------------------------------------
\begin{figure}[htbp]
\centering
\vspace{0.5cm}
	\caption{Isometric surface of the spin density 
$\rho_\uparrow-\rho_\downarrow$ $=\pm0.01$ states/(atomic unit)$^3$ 
in the energy range  $E_{\rm F}-0.6~{\rm eV} \ \sim \ E_{\rm F}$ 
in a thin slab parallel to (001) plane.  
Yellow and blue surfaces indicate the plus and minus spin densities,
respectively. Black solid lines stand for Ni-O bonds and red lines unit cell.}
\label{Fig-NdNiO3-SD}
\end{figure}
%----------------------------------------------
 
 The e$_g$ band in the range  $E_{\rm F} -0.6~{\rm eV} \sim E_{\rm F}$ does not
show any orbital ordering. In fact, off-diagonal elements within
the e$_g$ subblock of the occupation matrix $\{n_{mm'}\}$ are zero 
and the diagonal elements are identical. 
The absence of the orbital ordering may be consistent with the fact
that the $[101]$ axis has three-fold rotational symmetry in the
present spin configuration, once the distortion is neglected. 
Due to this pseudo three-fold rotational symmetry, the basis orbitals of
E$_g$ representation of the trigonal group D$_{3d}$ is a good
basis set and those derived from e$_g$ orbitals are
$\varphi_{u+}=\frac{1}{\sqrt{2}}(\varphi_{3z^2-r^2}+{\rm i}
\varphi_{x^2-y^2})$ and
$\varphi_{u-}=\frac{1}{\sqrt{2}}(\varphi_{3z^2-r^2}-{\rm i}
\varphi_{x^2-y^2})$. 
Therefore, there is no difference between the occupancies of
$\varphi_{3z^2-r^2}$ and $\varphi_{x^2-y^2}$.

Figures \ref{Fig-NdNiO3-SD} is the spatial profiles 
of spin densities $\rho_\uparrow-\rho_\downarrow$   
in the energy range  $E_{\rm F}-0.6~{\rm eV} \sim E_{\rm F}$
($\sigma^*$ state). 
Only one spin component of O p orbitals on the FM bond is bridging
between two Ni e$_g$ orbitals, while both spin components on the AFM
bond couple with Ni e$_g$ orbitals of respective spins. Consequently,
oxygen ions on the AFM bond have more charge in this energy range than
oxygen on the FM bond. This is the realization of oxygen-site
charge-ordered state discussed by
Mizokawa {\it et al.} in the framework of the Hartree-Fock 
calculation.~\cite{S-C-O} 
However, the charge difference between oxygen sites of
Ni$^{\uparrow}$-$\stackrel{\mbox{\tiny FM}}{\rm O}$-Ni$^{\uparrow}$ and
Ni$^{\uparrow}$-$\stackrel{\mbox{\tiny AFM}}{\rm O}$-Ni$^{\downarrow}$ 
in $\sigma^*$ states is mostly compensated 
by  hybridized  $\sigma$ state at lower energies.
Besides, all oxygen ions have no local magnetic moment.

More significant results seen in Fig.~\ref{Fig-NdNiO3-SD} may be the 
p orbital ordering on oxygen sites. 
The magnetic space group is $P_ba$ and  
its unitary part 
is $Pa$. 
The unit cell and lattice primitive vectors are not identical to those 
in $Pbnm$. 
The spin density in Fig.~\ref{Fig-NdNiO3-SD} is consistent with this 
magnetic group  $P_ba$, neither higher nor lower than this. 
The symmetry lowering of the unitary part $Pa$ is the origin of the opening 
the band gap at $E_{\rm F}$ in the majority spin band, 
despite to the absence of the orbital ordering on Ni sites
({\it i.e.} symmetry driven band gap). 
Therefore, one can conclude that 
the origin of the insulator phase at low temperatures 
in NdNiO$_3$ is the characteristic spin density 
 on the oxygen sites  or, equivalently,  
the orbital ordering there.
The energy gap $E_g=0.11$~eV is due to the symmetry 
of  charge order. 
In fictitious ideal  cubic structure without distortion 
or tilting of NiO$_6$ octahedra, the system becomes metal 
whose valence and conduction bands touch at points with each other.~\cite{RNiO3-GroupT}
The structure of $\sigma^*$ bands is 
insensitive to the value of $U$. 
The value of the band gap  is unchanged down to $U=4$~eV. 
This is because the gap is driven by the symmetry.

%----------------------------------------------
%----------------------------------------------

YNiO$_3$:
One should expect larger distortion in YNiO$_3$ because 
 YNiO$_3$ is the typical system with the small  tolerance 
factor.~\cite{YNiO3,RT-CO}
There are two different crystallographic Ni sites, 
and the distances from each Ni ion to surrounding O ion 
are different by 3$\sim$4\%
from one type of Ni ion to the other type.~\cite{YNiO3}
Calculated self-consistent solution is that with the apparent charge
disproportionation and no orbital polarization. 
Since Ni$^{4+}$ site has no spin magnetic moment, 
the two spin configurations discussed in NdNiO$_3$ become identical 
and the spin configuration of YNiO$_3$ is uniquely determined.
The magnetic space group is $P_b2_1$ if the spins of Ni$^{4+}$ ions 
are non vanishing and, on the contrary, 
$P_b2_1/a$ if the spins of Ni$^{4+}$ ions are zero. 
The latter symmetry $P_b2_1/a$ is actually the case. 

Figure~\ref{Fig-YNiO3-PDOS} shows the projected density of states at
Ni ion sites. 
The system is insulator with a gap $E_g=1.03~$eV. 
The resultant magnetic moments for half of Ni ions are 
$\pm 1.5\mu_B$ within the atomic sphere of a radius
$2.52~{\rm au}$, namely divalent ions 
Ni$^{2+}$ (t$_{2g}^6$e$_g^2$), of larger NiO$_6$ octahedron  
and zero for another half of Ni ions within the atomic sphere of a radius
$2.51~{\rm au}$,
namely quadrivalent Ni$^{4+}$ (t$_{2g}^6$), of smaller octahedron.
The experimentally observed magnetic moments are $\pm 1.4\mu_B$ 
for Ni$^{2+}$ ions
and $\pm 0.7\mu_B$ for Ni$^{4+}$ ions.~\cite{YNiO3}
The discrepancy may be due to a possible non-collinear spin order. 
The number of states in the energy range
 $E_{\rm F}-0.47~{\rm eV} \sim  E_{\rm F}~{\rm eV}$ 
is 16 and dominant weight on Ni$^{2+}$. 
Because the number of Ni$^{2+}$
in the $2 \times 1 \times 2$ cell is 8, these states are assigned as
two e$_g$ states mainly on Ni$^{2+}$ and surrounding oxygens. 
The e$_g$ orbitals of Ni$^{4+}$ is lifted in the higher energy range 
($>E_{\rm F} +1~{\rm eV}$) without spin polarization.

A large amount of e$_g$ orbitals in Ni$^{2+}$  ions  
locates at  the bottom of the d bands 
, in the range $E_{\rm F}-6.7~{\rm eV} \sim E_{\rm F}-5.8~{\rm eV}$
for Ni$^{2+}$, and 
in the range $E_{\rm F}-5~{\rm eV} \sim E_{\rm F}-4~{\rm eV}$
for Ni$^{4+}$, 
stabilizes the bonding between Ni$^{4+}$ ions and oxygen ions. 
From these facts, one can establish  a model that deep  $\sigma$
molecular orbitals (one per both Ni$^{2+}$ and Ni$^{4+}$) stabilize the system, and  
other two e$_g$ states  ($\sigma^*$ states) per one Ni$^{2+}$ ion  
located  at $E_{\rm F}-0.47~{\rm eV} \sim E_{\rm F}$.

%------------------------------------------
\begin{figure}[htbp]
\centering
	\vspace{0.5cm}
	\caption{Projected density of
	states of (a) divalent Ni$^{2+}$ and (b) quadrivalent Ni$^{4+}$ in
YNiO$_3$.
The energy zeroth is set at the top of the occupied states $E_{\rm
F}$. 
The local coordinate axes are chosen in the same manner for those in 
Fig.~1.}
\label{Fig-YNiO3-PDOS}
\end{figure}
%------------------------------------------

The charge disproportionation is mainly due to the crystal field
effect. 
The low spin state in Ni$^{4+}$ or Ni$^{3+}$ ion 
is energetically unstable in small $Dq$ case and 
 the ground state multiplet 
of Ni$^{2+}$ is $^{3}$A$_{2}$ (t$_{2g}^6$e$_g^2$) 
for all arbitrary values of $Dq$.~\cite{Tanabe}
Therefore, the small tolerance factor causes two different 
Ni sites, compressed Ni$^{4+}$ (large $Dq$) 
and dilated Ni$^{2+}$ (small $Dq$),  
rather than uniformly dilated Ni$^{3+}$ ionic states.~\cite{10Dq}
The standard values of $Dq/B$ ($B$ is the Racah parameter) 
for Ni$^{2+}$ is presumably around 1.0. 
Once one estimate the crystal field effects from the 
Ni-O bond lengths $d_{\rm Ni-O}$, 
the difference of $10Dq$ on two Ni ion sites is presumably  about 20\%.

Two narrow  e$_g$ bands can be seen in the energy range 
$E_{\rm F}-0.47~{\rm eV} \sim E_{\rm F}-0.23~{\rm eV}$ 
and $E_{\rm F}-0.23~{\rm eV} \sim E_{\rm F}$
in Fig.~\ref{Fig-YNiO3-PDOS}a. Two Ni$^{2+}$~e$_g$ states are spatially
extending over wide area and  not only hybridizing with the nearest
neighbor O ions but also extending over the nearest ${\rm Ni^{4+}}$ ions.  
This situation is well depicted in the spin density.
Figure~\ref{Fig-YNiO3-SD} shows the isometric surfaces of the spin
density in the range of 
$E_{\rm F}-0.47~{\rm eV} \sim E_{\rm F}-0.23~{\rm eV}$.
The d-wavefunctions on Ni$^{2+}$ are  extending over the (101) plane, 
and antiferromagnetically coupled with other Ni$^{2+}$ ions. 
Charge in each Ni ion is compensated in YNiO$_3$ as in NdNiO$_3$ 
and the difference between total charges in the muffin tin
spheres on Ni$^{2+}$ and Ni$^{4+}$ sites is very small, equals to 0.03. 
This variation is the same order as that of oxygen ions in NdNiO$_3$.
However, charge disproportionation is coupled with 
the lattice distortion, where larger oxygen octahedron is surrounding
Ni$^{2+}$, and stabilizes the lattice system
in YNiO$_3$. 
Therefore, diffraction experiment can detect the charge ordering
in the yttrium system easier than in the neodymium system.

%------------------------------------------
\begin{figure}[h]
\centering
	\vspace{0.5cm}
	\caption{ Isometric surface of spin density
    $\rho_\uparrow-\rho_\downarrow$ $=\pm0.002$~states/(atomic unit)$^3$
    in the energy range of an e$_g$ band ($E_{\rm F}-0.47$~eV$\sim$
$E_{\rm F}-0.23$~eV) in a thin slab parallel to (101) plane.
    Yellow surface means the sign of the spin density is plus,
    blue one  means minus.}
\label{Fig-YNiO3-SD}
\end{figure}
%------------------------------------------

%-----------------------------------------------------------------%
%\section{Conclusion}%

In conclusion, we have studied two typical antiferromagnetic
insulating phase of RNiO$_3$, NdNiO$_3$ and YNiO$_3$, by using the
LSDA+U method. 
A possibility of the Jahn-Teller distortion is excluded by the group
theoretical consideration. 
No orbital order on Ni sites is observed in both
systems, but two different types of ordering are observed.  In small
distorting RNiO$_3$ such as NdNiO$_3$, oxygen sites shows the orbital
ordering and this is the origin of the gap opening in NdNiO$_3$.  In
large distorting RNiO$_3$ such as YNiO$_3$, the charge
disproportionation 2Ni$^{3+}\rightarrow$Ni$^{2+}+$Ni$^{4+}$ occurs.
The charge ordering mechanism can explain the stabilization of the
asymmetric alignment of the local magnetic moments around each nickel
site  in both systems.

We may add the final comment about effects of electron-electron  correlation in RNiO$_3$.
The widths of the calculated  e$_g$ bands $D$ by the LDA are  
$2.82~{\rm eV}$(YNiO$_3$), 
$2.87~{\rm eV}$(NdNiO$_3$), and  
$3.18~{\rm eV}$(LaNiO$_3$).
Then, the ratio of $U/D$ may be estimated as 
$1:0.98:0.89$, assuming the value of the Coulomb repulsion $U$ is common 
for all,  and one can see a large reduction of $U/D$ in LaNiO$_3$, 
which may be the key parameter for the difference of the ground states 
of these perovskite nickelates.
NdNiO$_3$ shows the anomalous M-I transition.~\cite{NdNiO3-el}
LaNiO$_3$ is presumably an anomalous metal 
of strongly correlated electrons~\cite{Zhou}   
and could not be treated within the framework of the LSDA+U method.

%-----------------------------------------------------------------%
\section*{Acknowledgment}
The authors are very much grateful to V. I. Anisimov
for helpful advise on the LSDA+U method and for a fruitful discussion 
on the spin alignment of NdNiO$_3$, and also to  T. Mizokawa 
for useful discussion about their  work~\cite{S-C-O}. 
This work is supported by
Grant-in-Aid for COE Research ``Spin-Charge-Photon''.
%-----------------------------------------------------------------%

%-----------------------------------------------------------------%

\end{multicols}

\begin{references}
\bibitem{10Dq} The value of $10Dq$, the parameter of crystal field, is proportional 
to $d_{\rm Ni-O}^{-5}$.
\end{references}
\end{document}